\documentclass[twocolumn,aps,pra,superscriptaddress,showpacs,tightenlines]{revtex4}
\usepackage{amsmath}
\usepackage{amsfonts}
\usepackage{graphicx}
\usepackage{epsfig}
\usepackage{color}
\usepackage[colorlinks,citecolor=blue]{hyperref}
\begin{document}

\title{Collective Radiance Effects in the Ultrastrong Coupling Regime}
\author{Qian Bin}
\affiliation{School of Physics, Huazhong University of Science and Technology, Wuhan, 430074, P. R. China}
\author{Xin-You L\"{u}}
\email{xinyoulu@hust.edu.cn}
\affiliation{School of Physics, Huazhong University of Science and Technology, Wuhan, 430074, P. R. China}
\author{Tai-Shuang Yin}
\affiliation{School of Physics, Huazhong University of Science and Technology, Wuhan, 430074, P. R. China}
\author{Yong Li}
\affiliation{Beijing Computational Science Research Center, Beijing 100193, China}
\author{Ying Wu}\email{yingwu2@126.com}
\affiliation{School of Physics, Huazhong University of Science and Technology, Wuhan, 430074, P. R. China}
\date{\today}
\begin{abstract}
We investigate the collective radiance characteristics of qubits in the ultrastrong coupling regime, where the radiance witness is defined based on the resonator-qubit dressed basis. The ultrastrong hyperradiance effect is demonstrated when the dressed state of system is resonantly driven. Interestingly, we show that, besides the resonator-qubit coupling strength, the parity-symmetry-breaking induced cascade-transition can significantly enhance the collective radiance of qubits, which allows us to manipulate the transitions between subradiance, superradiance, and hyperradiance via adjusting the parity symmetry of system with an external magnetic field.
This work extends the collective radiance theory to the ultrastrong coupling regime, and offers the potential applications in the engineering of new laser devices.
\end{abstract}
\maketitle
\section{introduction}
The theory of the collective radiance is of great importance in quantum optics, and has important applications in lasing engineering~\cite{A. A. Svidzinsky,J. G. Bohnet,Q. Baudouin}, precision measurement~\cite{W. -J. Kim,R. Rohlsberger,M. A. Norcia}, and quantum information~\cite{A. Kuzmich,A. Asenjo-Garcia}. One of the very intriguing phenomena exhibiting the collective radiance behavior is superradiance discovered by Dicke in 1954~\cite{R. H. Dicke}. Specifically, the radiance intensity from an atomic ensemble can be enhanced with a factor of $N^2$ ($N$ is the atom number). Recently, the enhanced radiance factor that is larger than $N^2$ has also been present in the cavity quantum electrodynamics (QED) system, called as hyperradiance~\cite{M. -O. Pleinert,J. P. Xu,Y. F. Han}, which has stronger collective radiance effect than superradiance. The above theoretical results push the corresponding experimental progresses, including superradiance laser~\cite{J. G. Bohnet,M. A. Norcia2}, the measurement of  collective Lamb Shift~\cite{R. Rohlsberger,J. Keaveney}, the research of coherence properties of Bose-Einstein Condensate~\cite{S. Inouye,D. Schneble}, and the realization of superradiance in quantum dots~\cite{M. Scheibner} and artificial atomic systems~\cite{J. A. Mlynek}. However, the present collective radiance theories are confined to the weak and strong coupling regimes.

Recently, ultrastrong coupling regime, the light-matter coupling rate reaching the order of $10\%$ of the bare resonance frequency of photons or the transition frequency of quantum emitters, has been reached experimentally in a variety of solid state quantum systems~\cite{G. Gunter,T. Niemczyk,Y. Todorov,P. Forn-Diaz,T. Schwartz,A. J. Hoffman,C. M. Wilson,G. Scalari,S. Gambino,P. Forn-Diaz2}.  In this regime, the counter-rotating terms in the interaction Hamiltonian are non-ignorable~\cite{C. Ciuti}, and in some cases, the parity symmetry of system can not be conserved approximately~\cite{F. Deppe,A. Fedorov}. Thus many novel quantum effects emerge in the ultrastrong coupling regime, such as vacuum degeneracy~\cite{P. Nataf}, the generation of correlated photon pairs from the initial polariton vacuum state~\cite{C. Ciuti2}, non-classical state~\cite{S. Ashhab}, Casimir-like photon~\cite{F. Beaudoin} and so on~\cite{Z. H. Wang,A. P. Hines,E. K. Irish,J. Larson,J. Bourassa,D. Zueco,C. P. Meaney,I. Lizuain,
A. Ridolfo,R. Stassi,C. Maissen,E. Sanchez-Burillo,G. Scalari2,K. K. W. Ma,L. Garziano,Q. Bin,P. Nataf2}. Then extending the collective radiance theory to the ultrastrong coupling regime becomes interesting in the exploration of  novel effects and the application of the lasing theory.

In the ultrastrong coupling regime, the usual radiance witness~\cite{M. -O. Pleinert,J. P. Xu,Y. F. Han}, defined by an independent system operator, is failed for the description of collective radiance characteristics of system. Here, we rewrite the radiance witness in terms of resonator-qubit dressed basis, which is valid for any resonator-qubit coupling strength. We find that the ultrastrong resonator-qubit coupling could significantly enhance the collective radiance of qubits under the condition of resonantly driving the dressed state of system, which leads to the emergence of enhanced hyperradiance. The parameter ranges for different radiance effects, such as subradiance and hyperradiance, become more distinguishable. Moreover, we also show that the resonator-qubit detuning could change the property of the radiance, allowing for transitions between different radiance effects.

More interestingly, when the coupling strength is fixed, the collective radiance of qubits can also be enhanced by breaking the parity symmetry of system, e.g., making the radiance property of system from subradiance and superradiance to hyperradiance. This originally comes from the symmetry-breaking-induced cascade-transition of decay during the dressed states. Note that, in some cases (e.g., the superconducting circuits), the system parity symmetry could be controlled by an external magnetic field~\cite{F. Deppe,Y. X. Liu}. Then our results allow us to realize the controllable transitions between the subradiance, superradiance and hyperradiance via adjusting the system parity symmetry, which might inspire new laser technologies. Our work is also fundamentally interesting in building the collective radiance theory in the ultrastrong coupling regime.

\section{System and collective radiance witness.}
\begin{figure}
\includegraphics[width=8.5cm]{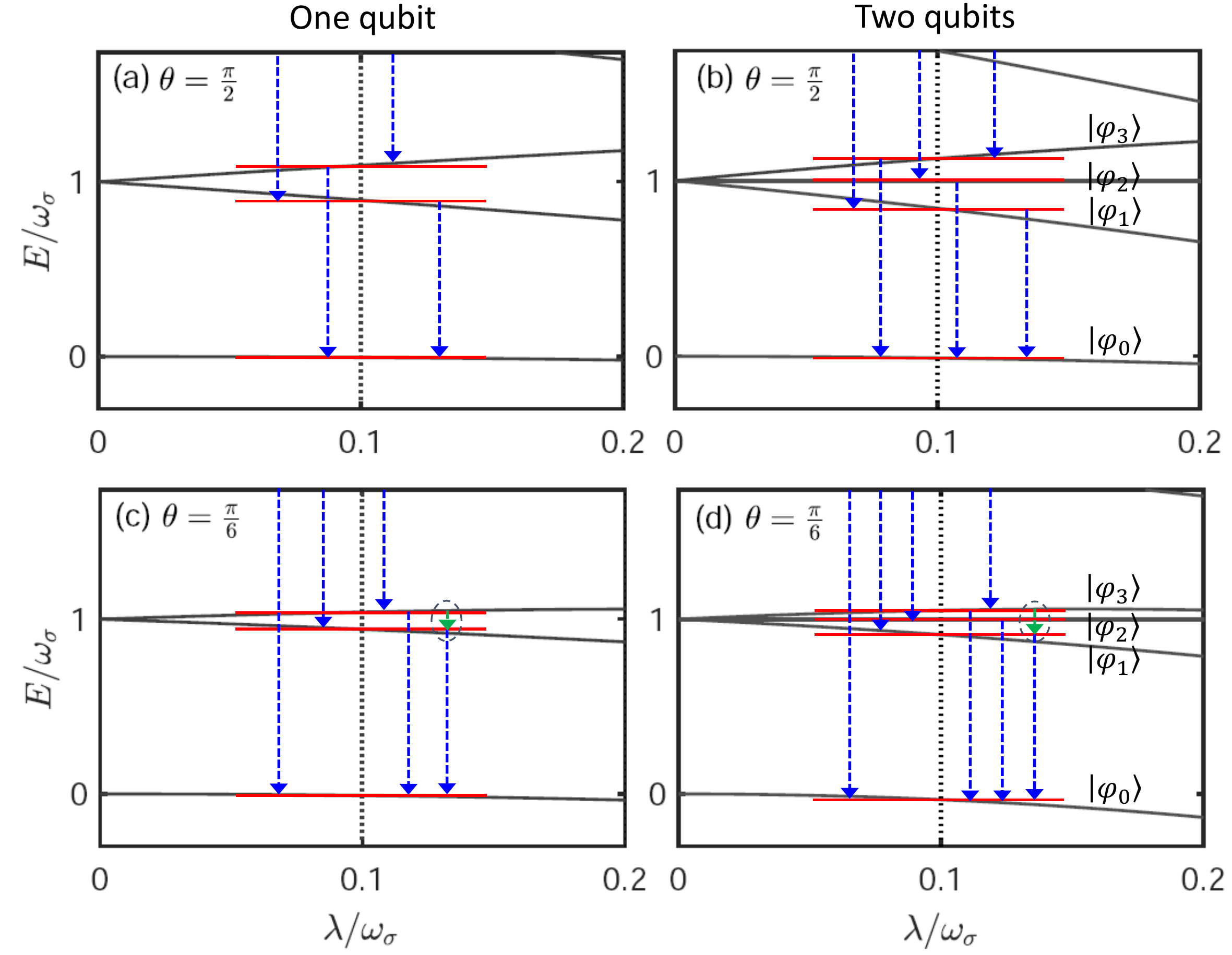}\\
\caption{ Energy spectrum of Hamiltonian $H_0$ versus the coupling strength $\lambda/\omega_{\sigma}$ for (a,b) $\theta=\pi/2$ and (c,b) $\theta=\pi/6$. Moreover, the subplots (a,c) and (b,d) correspond to the case of one qubit and two qubits, respectively. Here $|\varphi_n\rangle$ $(n=0,1,2...)$ are the corresponding eigenstates of $H_0$. The arrows show the possible transitions of radiative decay between these eigenstates. Especially, the green arrow indicates the symmetry-breaking-induced cascade-transition of decay. Here the system parameters are chosen as $\omega_c=\omega_{\sigma}$ and $\lambda/\omega_{\sigma}=0.1$.}\label{fig1}
\end{figure}

We consider a circuit-QED system that consists of a single-mode resonator coupled to two qubits driven by coherent microwave field. The Hamiltonian of system can be given by ($\hbar=1$)
\begin{equation}\label{eq1}
 H=H_0+H_d
\end{equation}
with
\begin{align}\label{eq2}
H_0=&\omega_c a^\dag a+\omega_{\sigma}\sum_j\sigma^+_j\sigma_j^-\\\nonumber
&+\lambda(a^\dag+a)\sum_{j=1}^2(\cos\theta\sigma_z^j -\sin\theta\sigma_x^j),
\end{align}
where, $a$ ($a^\dag$) is the annihilation (creation) operator of the resonator with resonance frequency $\omega_c$,  $\sigma_j^+$ ($\sigma_j^-$) is the raising (lowering) operator for the \emph{j}th qubit with transition frequency $\omega_{\sigma}$, and $\sigma_x^j=\sigma_j^-+\sigma_j^+$, $\sigma_z^j=\sigma_j^+\sigma_j^--\sigma_j^-\sigma_j^+$, The constant $\lambda$ is the coupling rate between the resonator mode and each qubit. The term $H_d=\Omega\cos(\omega_d t)\sum^2_{j=1}(\sigma_j+\sigma^+_j)$ describes the coherent driving with driving frequency $\omega_d$ and amplitude $\Omega$.

Here, the mixing angle $\theta$, describing the relative contribution of the transverse and longitudinal couplings, can be controlled by adjusting an external magnetic flux $\Phi_{\rm x}$ threading the qubit loop, i.e., $\sin \theta=\Delta/\omega_{\sigma}$, where $\Delta$ is the qubit energy gap~\cite{F. Deppe,Y. X. Liu,R. Stassi2}. The value of $\theta$ can influence the transition of radiance via changing the parity symmetry of the system, and will further produce an effect on the collective radiance property of the system. Here, the parity operator of the system is defined as $\Pi=\exp[i\pi N]=\exp[i\pi(a^\dag a+\sigma_1^+ \sigma_1^-+\sigma_2^+ \sigma_2^-)]$~\cite{J. Casanova,D. Braak,M. J. Hwang,L. -L. Zheng}. For $\theta=\pi/2$, with $[H_0,\Pi]=0$, the parity of the number of excitations in the Hamiltonian $H_0$ is conserved. However, the parity-symmetry of $H_0$ is broken when $\theta\neq\pi/2$, i.e., $[H_0,\Pi]\neq0$. This enables a cascade transition between adjacent dressed states, e.g., the transition $|\varphi_3\rangle\rightarrow|\varphi_1\rangle$ in Fig.\,\ref{fig1}(d), which is forbidden for the case of parity symmetry conservation. Here, the dressed states are given approximately, i.e., $|\varphi_1\rangle\approx(|e,g,0\rangle+|g,e,0\rangle)/2+|g,g,1\rangle/\sqrt{2}$ and $|\varphi_3\rangle\approx(|e,g,0\rangle+|g,e,0\rangle)/2-|g,g,1\rangle/\sqrt{2}$. Note that the term leading parity-symmetry-breaking can be ignored safely by the rotating wave approximation (RWA) in the weak and strong coupling regimes.

To describe the system more realistically, the influence of dissipation on the system needs to be taken into account. The system coupled to a zero temperature environment can be studied by a quantum optical master equation. However, the standard master equation is failed to provide a correct description for the dynamics of the system in the case of $\lambda\sim\omega_c,\omega_{\sigma}$. Because in the ultrastrong coupling regime, the qubits and resonator mode can form an inseparable system with the new dressed states. We thus write the system Hamiltonian operators in terms of the resonator-qubit dressed basis $|\varphi_n\rangle$ $(n=0,1,2...)$, where $H_0|\varphi_n\rangle=E_n|\varphi_n\rangle$. By applying Born-Markov approximation and tracing out the environment degrees of freedom, the master equation for the reduced density matrix of the system reads~\cite{A. Ridolfo,R. Stassi,L. Garziano}
\begin{align}\label{eq3}
\frac {d\rho}{dt}=&i[\rho,H]+\kappa\mathcal{L}[X^+]
+\gamma_{\sigma}\sum_{j=1}^N\mathcal{L}[D^+_j],
\end{align}
where the Liouvillian superoperator $\mathcal{L}$ is defined as $\mathcal{L}[\emph{O}]=(2\emph{O}\rho\emph{O}^\dag-\rho\emph{O}^\dag\emph{O}-\emph{O}^\dag\emph{O}\rho)/2$. The constants $\kappa$ and $\gamma_{\sigma}$ describe the damping rates of the cavity and the qubits, respectively. Here,  $X^+=\sum_{E_n,E_m>E_n}X_{nm}|\varphi_n\rangle \langle\varphi_m|$ and $D^+_j=\sum_{E_n,E_m>E_n}D_{nm}^j|\varphi_n\rangle \langle\varphi_m|$, with $X_{nm}=\langle \varphi_n|(a+a^\dag)|\varphi_m\rangle$ and $D_{nm}^j=\langle \varphi_n|(\sigma_j^-+\sigma^+_j)|\varphi_m\rangle$,  are positive frequency components of the cavity photon and the \emph{j}th qubit operators, respectively. Note that $a|\varphi_0\rangle\neq0$ for the ground state of the Hamiltonian $H_0$, and $X^+|\varphi_0\rangle=0$. Under the condition of including the RWA or neglecting the resonator-qubit coupling rate, $X^+$ and $X^-=(X^+)^\dag$ correspond approximately to $a$ and $a^\dag$. Similarly, $D^+_j$ and $D_j^-=(D^+_j)^\dag$ coincide with $\sigma_j^-$ and $\sigma_j^+$.

In this work, we explore the collective radiance characteristics of qubits in the steady-state limit. According to the input-output theory, the photon emission from qubits can be measured by detecting the average photon number from cavity. The output photon rate of the cavity is expressed as $\Psi_{\rm out}=\kappa\langle X^-X^+\rangle$, obtained by the input-output relation $a_{\rm out}(t)=a_{\rm in}(t)-\sqrt{\kappa}X^+(t)$ in the case of $\omega_c\approx\omega_{\sigma}$, where the input is in the vacuum. The photon emission could be detected in photodetection experiment by coupling the qubit to a microwave antenna~\cite{M. Hofheinz}. The radiance characteristics of two qubits can be described by a radiance witness
\begin{align}\label{eq4}
R=\frac{\langle X^-X^+ \rangle_2-2\langle X^-X^+\rangle_1}{2\langle X^-X^+\rangle_1}.
\end{align}
Here, $\langle X^-X^+ \rangle_2$ is the average photon number when a cavity is coupled to two qubits, and $\langle X^-X^+ \rangle_1$ corresponds to the case of coupling the cavity to only one qubit. Under this definition, $R=0$ indicates an uncorrelated radiance between two qubits. Specifically, the emission photons of two qubits are the sum of that of two isolated qubits, i.e., $\langle X^-X^+ \rangle_2=2\langle X^-X^+ \rangle_1$. $R<0$ corresponds to the subradiance of two qubits, i.e., $\langle X^-X^+ \rangle_2<2\langle X^-X^+ \rangle_1$, indicating the suppression of radiance. The range of $0<R\leq1$ corresponds to the regime of superradiance, and $R=1$ means that the radiance strength being proportional to the square of the number of qubits, i.e., $\langle X^-X^+ \rangle_2=2^2\langle X^-X^+ \rangle_1$. $R>1$, i.e., $\langle X^-X^+ \rangle_2>2^2\langle X^-X^+ \rangle_1$, is the hyperradiance, which has stronger radiance effect than the superradiance behavior~\cite{M. -O. Pleinert,J. P. Xu,Y. F. Han}.

\section{Radiance without parity symmetry breaking}
\begin{figure}
  \centering
  \includegraphics[width=8.5cm]{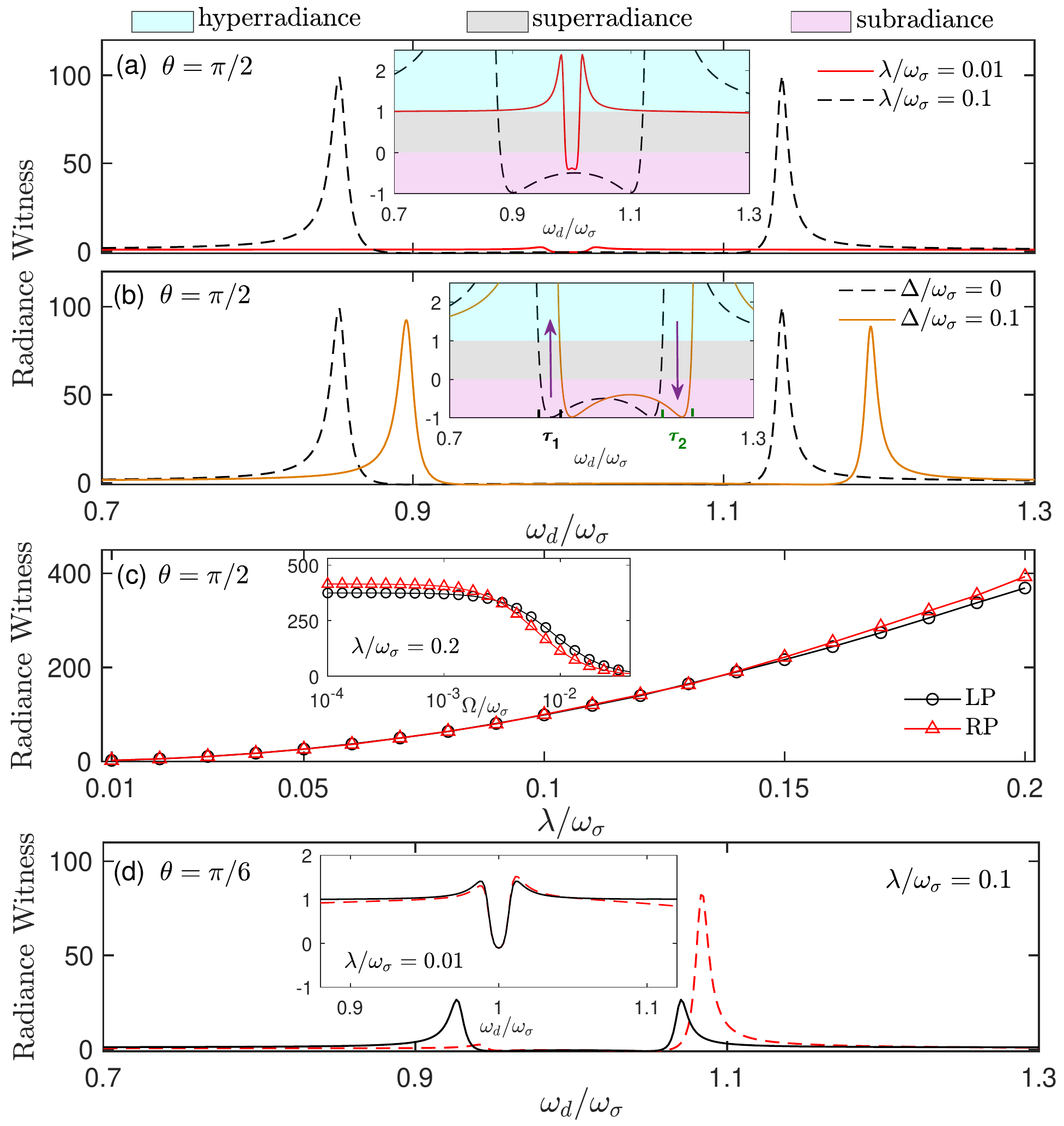}\\
  \caption{ Radiance witness $R$ versus $\omega_d/\omega_{\sigma}$ for different (a) $\lambda/\omega_{\sigma}$ and (b) $\Delta/\omega_{\sigma}$. Insets: the enlarged region of the small value of $R$. The blue, gray and pink areas indicate $R>1$ (hyperradiance), $0<R\leq1$ (superradiance) and $-1<R<0$ (subradiance), respectively. Here $\tau_1\in(0.875,0.917)$ and $\tau_2\in(1.12,1.178)$. (c) The values of left peak (LP) and right peak (RP) in (a) versus $\lambda/\omega_{\sigma}$ and $\Omega/\omega_{\sigma}$ (the inset). (d) $R$ versus $\omega_d/\omega_{\sigma}$ for different $\lambda/\omega_{\sigma}$ when the term including $\sigma_z^j$ is neglected (black solid lines) and kept (red dashed lines). The other system parameters used here are: $\gamma/\omega_{\sigma}=\kappa/\omega_{\sigma}=0.01$, $\Omega/\omega_{\sigma}=0.001$, and (a,c,d) $\omega_c=\omega_{\sigma}$, (b) $\lambda/\omega_{\sigma}=0.1$.}\label{fig2}
\end{figure}

To clearly show the influence of resonator-qubit coupling strength on the radiance effect of qubits, we investigate the case $\theta=\pi/2$ (holding the parity symmetry of system) in Figs.\,\ref{fig2}(a)-\ref{fig2}(c). It shows that the collective radiance effect of the qubits is significantly influenced in the ultrastrong coupling regime. Under different driving frequencies, one can obtain the subradiance, superradiance and hyperradiance, respectively. For example, when we resonantly drive the dressed state $|\varphi_{1}\rangle$ (or $|\varphi_{3}\rangle$) of the system including two qubits, the strong hyperradiance is obtained, which corresponds to the peaks in Fig.\,\ref{fig2}(a). This comes from the emission of photons between the two states $|\varphi_{1}\rangle$ (or $|\varphi_{3}\rangle$) and $|\varphi_{0}\rangle$ with different transition paths~\cite{M. -O. Pleinert}. For the case of resonantly driving the dressed state of the system that consisting of one qubit, the subradiance can be obtained, corresponding to the deeps in Fig.\,\ref{fig2}(a). These imply that there are different optimal radiance frequencies for the systems containing different numbers of qubit, as shown in Figs.\,\ref{fig1}(a) and \ref{fig1}(b).

\begin{figure}
  \centering
  \includegraphics[width=8.6cm]{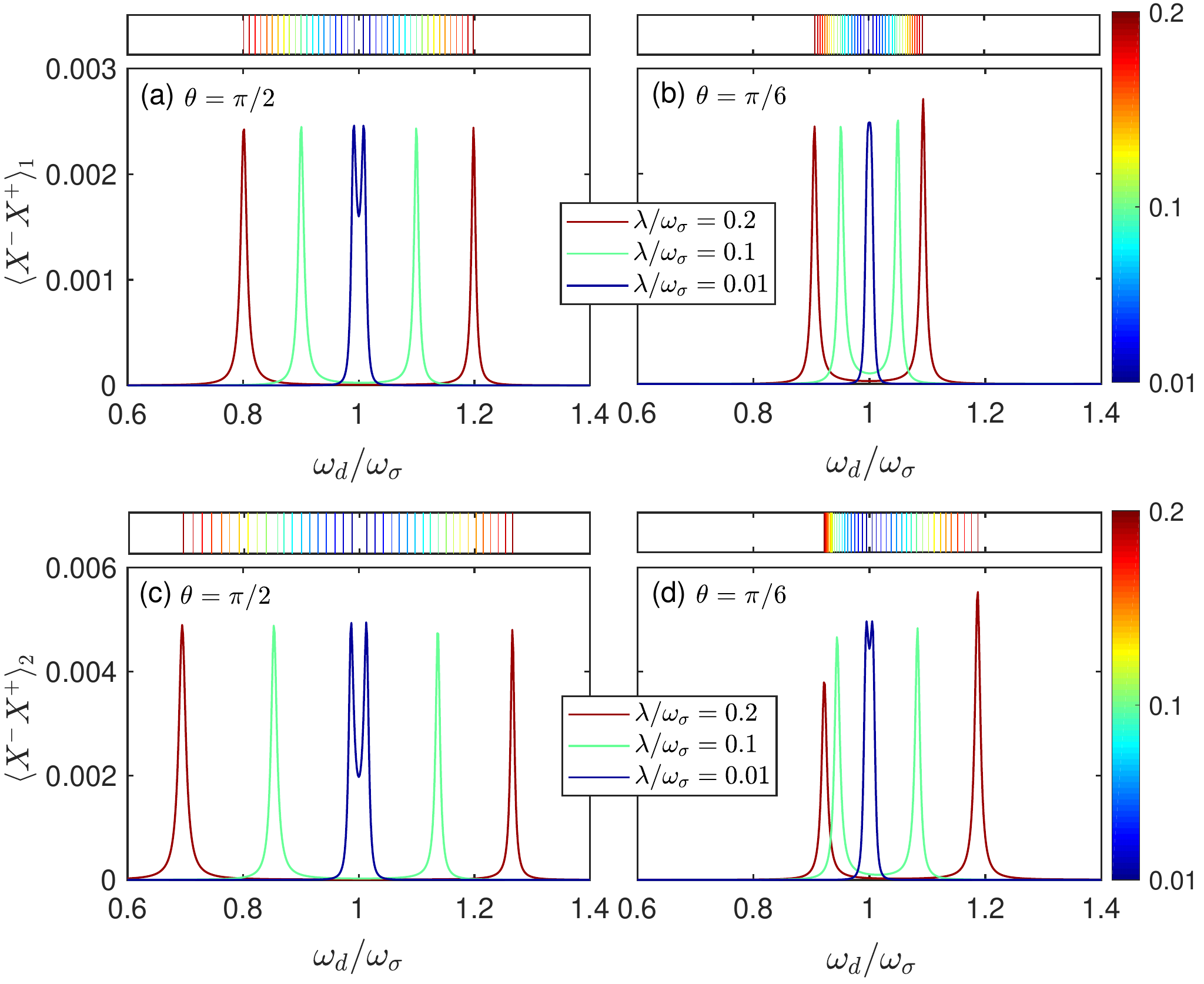}\\
  \caption{The excitation spectrum of the systems consisting of one qubit (a,b) and two qubits (c,d). Here we present the change of excitation spectrum with the increasing of $\lambda/\omega_{\sigma}$, and the top stripes indicate the positions of peaks for different $\lambda/\omega_{\sigma}$. The other system parameters used here are: $\omega_c=\omega_{\sigma}$, $\gamma/\omega_{\sigma}=\kappa/\omega_{\sigma}=0.01$ and $\Omega/\omega_{\sigma}=0.001$. }\label{fig4}
\end{figure}

The distance between the peak and the deep is getting farther and farther away with increasing the resonator-qubit coupling strength $\lambda$. This result can be understood from the energy spectrum [see Figs.\,\ref{fig1}(a) and \ref{fig1}(b)] and the corresponding excitation spectrum [see Figs.\,\ref{fig4}(a) and \ref{fig4}(c)]. From Figs.\,\ref{fig4}(a) and \ref{fig4}(c), we also see that, as increasing the resonator-qubit coupling strength, the system consisting of a resonator coupled two qubits has a faster splitting speed between two adjacent dressed states than that coupled one qubit. In other words, when we fix the value of $\lambda$, the splitting between two peaks is larger than that of two deeps, which leads to an increase in the distance between the peak and the deep. Then, in the ultrastrong coupling regime, the regions of different radiance effects (e.g., superradiance and hyperradiance) become more distinguishable.

In Fig.~\ref{fig2}(c), we plot the dependence of the maximum radiance strength on $\lambda$ and $\Omega$. It also shows that the collective radiance effect is enhanced as increasing the value of $\lambda$, whereas too large drive strength will decrease the collective radiance of the qubits. This is because it is difficult to neglect the higher-order dressed states when $\Omega$ has a higher value. So the possible radiative transitions could lead to the occurrence of the destructive quantum path interference in the system. Moreover, we also show the influence of resonator-qubit detuning $\Delta$ ($\Delta=\omega_c-\omega_{\sigma}$) on the radiance effect of qubits in Fig.~\ref{fig2}(b). In the presence of detuning , the enhanced collective radiance effects are still persisting, but the curve has an obvious shift due to the shift of dressed states. This allows for transitions between subradiance, superradiance and hyperradiance within the proper parameter ranges [see the arrow in the ranges $0.875<\omega_d/\omega_{\sigma}<0.917$ and $1.12<\omega_d/\omega_{\sigma}<1.178$ in Fig.\,\ref{fig2}(b)].

Even for the case $\theta\neq\pi/2$, the system can also hold the parity symmetry in the weak coupling regime when we ignore the term $\lambda(a^\dag+a)\sum_{j=1}^2\cos\theta \sigma_z^j$ under RWA. However, in the ultrastrong coupling regime, the RWA becomes invalid and the parity symmetry of the system is broken. This property is clearly shown in Fig.\,\ref{fig2}(d), which indicates that the system parity symmetry can significantly influence the collective radiance.
\section{Radiance with parity symmetry breaking}
\begin{figure}
  \centering
  \includegraphics[width=8.6cm]{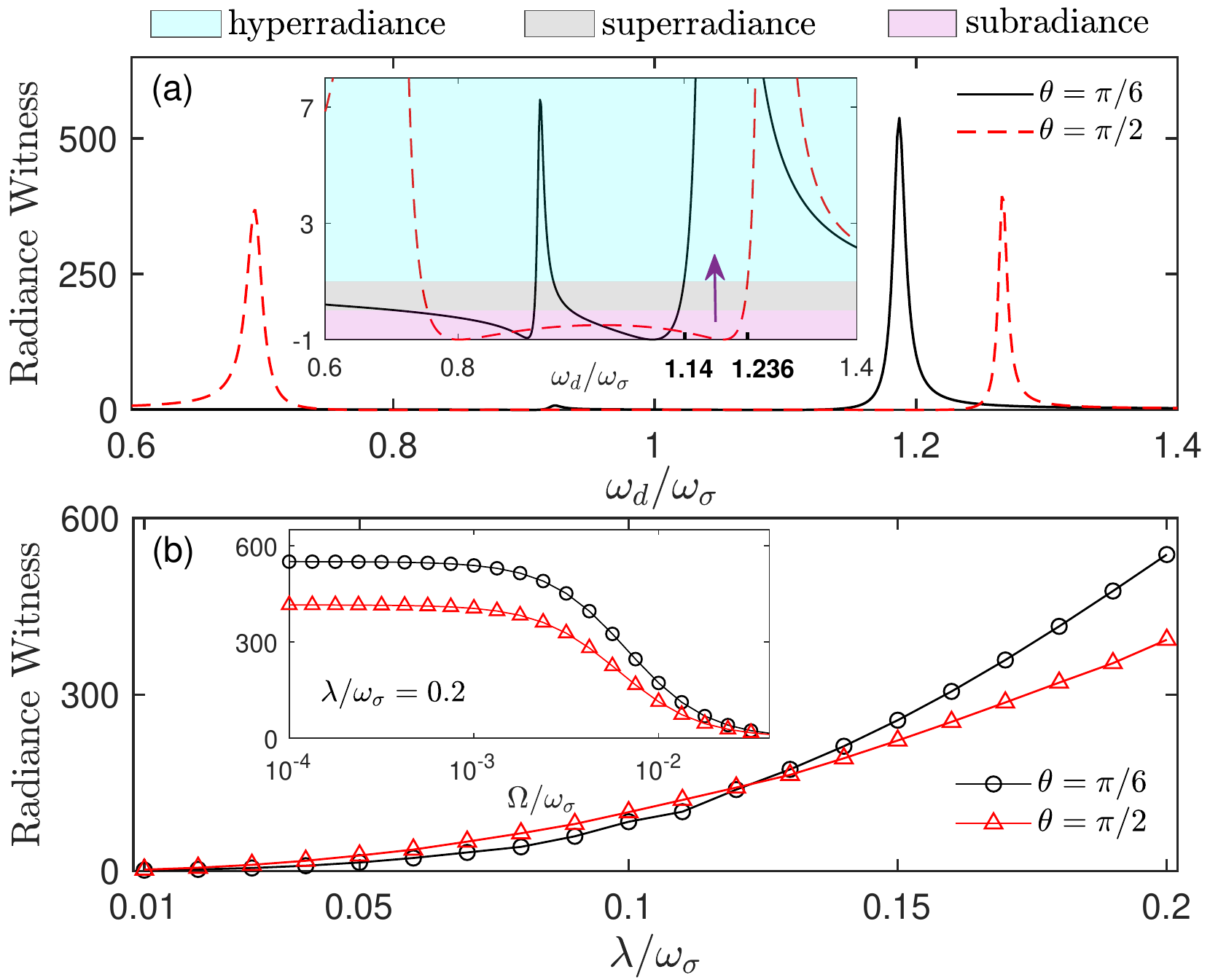}\\
  \caption{ (a) Radiance witness $R$ versus $\omega_d/\omega_{\sigma}$ for different $\theta$ when $\lambda/\omega_{\sigma}=0.2$. Inset: the enlarged region of the small value of $R$. The blue, gray and pink areas indicate $R>1$ (hyperradiance), $0<R\leq1$ (superradiance) and $-1<R<0$ (subradiance), respectively. (b) The maximum value of $R$ in Fig.\,\ref{fig3}(a) and Fig.\,\ref{fig2}(a) versus $\lambda/\omega_{\sigma}$ and $\Omega/\omega_{\sigma}$ (the inset). The other system parameters used here are: $\omega_c=\omega_{\sigma}$, $\gamma/\omega_{\sigma}=\kappa/\omega_{\sigma}=0.01$ and $\Omega/\omega_{\sigma}=0.001$ .}\label{fig3}
\end{figure}
Now let's investigate the influence of parity symmetry of system on the collective radiance effect in detail.

Firstly, in Fig.\,\ref{fig3}(a), we plot the radiance witness $R$ as a function of $\omega_d/\omega_{\sigma}$ when $\theta=\pi/2$ (holding parity symmetry) and $\theta=\pi/6$ (breaking parity symmetry).
It is shown that the parity-symmetry-breaking will significantly enhance the collective radiance when one resonantly drives the upper dressed state $|\varphi_3\rangle$. Physically, this enhancement comes from the cascade-transition of decay during the dressed states, i.e, $|\varphi_3\rangle\rightarrow|\varphi_1\rangle$, induced by the counter-rotating terms $a^\dag\sigma_z^j$ and $a\sigma_z^j$. Specifically, when the upper dressed state $|\varphi_3\rangle$ is resonantly excited, besides the radiance transition $|\varphi_3\rangle\rightarrow|\varphi_0\rangle$, the parity-symmetry-breaking induced radiance transitions $|\varphi_3\rangle\rightarrow|\varphi_1\rangle\rightarrow|\varphi_0\rangle$ will also emit photons, which significantly enhance the radiance effect of the qubits. However, this cascade-radiance transition does not exist in the case of driving the lower dressed state $|\varphi_1\rangle$. Together with the reduced resonator-qubit interaction strength $\lambda\sin\theta$ for $\theta\neq\pi/2$, the radiance effect is suppressed by the parity-symmetry-breaking when we resonantly drive the lower dressed state $|\varphi_1\rangle$ [see the left peaks in Fig.\,\ref{fig3}(a)].

Secondly, from Fig.\,\ref{fig3}(a), we also see that the positions of the peaks and deeps have some shifts when the parity symmetry of the system is broken. This leads to a parity-symmetry-breaking induced transition from subradiance and superradiance to hyperradiance during the proper parameter range [see the arrow in the range $1.14<\omega_d/\omega_{\sigma}<1.263$ in Fig.\,\ref{fig3}(a)]. Then our results allow the realization of controllable radiance transition in the system with controllable parity symmetry. For example, in the superconducting circuit, one could obtain the transition from subradiance to hyperradiance by breaking system parity symmetry with an external magnetic field. To understand the above result, we plot the excitation spectrum of the systems including one qubit and two qubits, respectively, in Fig.\,\ref{fig4}. Note that the resonant excitation frequencies for the cases of one qubit and two qubits correspond to the positions of deeps and peaks, respectively, in Fig.~\ref{fig3}(a). Comparing the cases of $\theta=\pi/6$ and $\theta=\pi/2$, we see that the parity-symmetry-breaking destroys the symmetry of the excitation spectrum, which ultimately leads to the shifts of the peaks and deeps of $R$ in Fig.~\ref{fig3}(a).

Lastly, in Fig.~\ref{fig3}(b), we plot the dependence of the maximum radiance strength on $\lambda$ and $\Omega$ for $\theta=\pi/6$ and $\theta=\pi/2$.
It shows that the parity-symmetry-breaking can enhance the radiance effect in the ultrastrong coupling regime. Note that this enhancement effect can be ignored approximately in the weak and strong coupling regimes. This result is consistent with the Fig.\,\ref{fig2}(d). Physically, the parity-symmetry-breaking can enhance the radiance of the qubits by inducing the cascade-transition of decay ($|\varphi_3\rangle\rightarrow|\varphi_1\rangle\rightarrow|\varphi_0\rangle$). However the presence of the term $\sin\theta$ of Eq.~(\ref{eq2}) decreases the collective radiance effect by reducing the effective resonator-qubit coupling strength when $\theta\neq\pi/2$. The change of the radiance effect is the result of the competition between the symmetry-breaking-induced cascade-transition of decay and the decreasing of effective coupling strength. Thus, in Fig.~\ref{fig3}(b), we see that the maximum radiance strength in the parity-symmetry-breaking system is greater than that in the parity-symmetry-conserving system when $\lambda/\omega_{\sigma}<0.12$. For a weaker coupling strength, the effect from the decreasing of coupling strength is also larger than that of the cascade-transition in the parity-symmetry-breaking system.
\section{Conclusions and discussions}

We have investigated the influences of resonator-qubit coupling strength, resonator-qubit detuning and system parity symmetry on the collective radiance characteristics of circuit-QED system in the ultrastrong coupling regime. We have shown that, besides the ultrastrong coupling strength, the parity-symmetry-breaking will also enhance the collective radiance effect significantly by inducing the cascade-transition between two adjacent dressed states of system. Moreover, the resonator-qubit detuning and parity-symmetry-breaking of system will also shift the positions of subradiance, superradiance and hyperradiance largely. This result provides the potential methods to manipulate the transitions between subradiance, superradiance and hyperradiance via adjusting resonator-qubit detuning or system parity symmetry. Note that usually when referring to the superradiance, it is easy to associate it with superradiance phase transition. However, the model we are considering is a dynamic process, so the superradiance (and hyperradiance) phenomenon in our studying is different from the superradiance involving the ground state in the usual phase transition approach~\cite{K. Hepp,Y. K. Wang}.

In addition, we discuss that the superconducting circuit is an ideal experimental platform for our studying. A possible implementation is in the system that consists of a superconducting coplanar waveguide resonator galvanically coupled to two flux qubits threaded by the external flux bias. In this system, the mixing angle $\theta$ can be adjusted by the flux bias threading the qubit loop~\cite{T. Niemczyk,F. Deppe,A. Ridolfo}. In principle, our results are also fit for the acoustic system, and then the new type of photon and phonon laser devices might be inspired by our work.

In future works, it will be interesting to extend the collective radiance to the deep strong coupling regime. In this coupling regime, although it is not easy to calculate the collective radiance effect by numerical and analytical methods in the resonator-qubit resonance system, we can consider a solvable or quasi-solvable model of large detunings~\cite{J. Casanova,D. Braak,M. J. Hwang,M. Bina1,J. Peng,M. Bina2}. This will bring more interesting results to the study of collective radiance in the systems without RWA.

This work is supported by the National Key Research and Development Program of China grant
2016YFA0301203, the National Science Foundation of China (Grant Nos. 11374116, 11574104 and 11375067).

\end{document}